\documentclass[%
superscriptaddress,
twocolumn,
amsmath,amssymb,
aps,
prl,
]{revtex4-2}

\usepackage{graphicx}% Include figure files
\usepackage{dcolumn}% Align table columns on decimal point
\usepackage{bm}% bold math
\usepackage{xcolor} % for \textcolor
\usepackage{hyperref}
\usepackage{amsmath}
\usepackage{amssymb}
\usepackage{mathtools}
\usepackage{comment}
\usepackage{float}

\newcommand{\omp}{\omega_{\rm p}}
\newcommand{\omo}{\omega_{\rm 0}}

\newcommand{\omB}{\omega_{\rm B}}
\newcommand{\OmB}{\Omega_{\rm B}}
\newcommand{\omop}{\omega_{\rm 0}/\omega_{\rm p}}

\newcommand{\varepsilonB}{\varepsilon_{\rm B}}
\newcommand{\varepsilonp}{\varepsilon_{\rm p}}
\newcommand{\ahat}{\hat{a}}

\newcommand{\me}{m_{\rm e}}

\newcommand{\eqb}{\begin{eqnarray}}
\newcommand{\eqe}{\end{eqnarray}}
\newcommand{\be}{\begin{eqnarray}}
\newcommand{\ee}{\end{eqnarray}}
\newcommand{\bi}{\begin{itemize}}
\newcommand{\ei}{\end{itemize}}

\newcommand{\llinear}{\ell_{\rm lin}}

\begin{document}

% \preprint{APS/123-QED}

\title{Interaction of Strong X-mode Waves with Magnetized Pair Plasmas}

\author{Navin Sridhar}
\email{navinsridhar@stanford.edu}
\affiliation{Department of Physics, Stanford University, 382 Via Pueblo Mall, Stanford, CA 94305, USA} 
\affiliation{Kavli Institute for Particle Astrophysics \& Cosmology, 452 Lomita Mall, Stanford, CA 94305, USA}

\date{\today}% It is always \today, today,

\begin{abstract}
The interaction of strong ($a_0>1$), superluminal X-mode electromagnetic waves with magnetized pair plasma falls into two regimes set by $\varepsilonB\equiv a_0\omB/\omega_0$, the ratio of the wave strength parameter to its frequency, in units of cyclotron frequency. For $\varepsilonB<1$ the wave propagates until induced scattering disrupts it. Using analytical theory and kinetic simulations, we show that the background magnetic field introduces a restoring frequency $\OmB$ to scattering; when $\OmB$ exceeds the scattering rate, the unattenuated length grows from $\simeq\varepsilonp^{-2/3}$ to $\simeq\varepsilonB^{1/2}/\varepsilonp$ wavelengths ($\varepsilonp\equiv a_0\omega_{\rm p}/\omega_0$; $\omp$ is the plasma frequency). When $a_0^2>\varepsilonB>1$, the plasma current is nonlinear, and the wave is damped while riding a motional electric field, so one half-cycle appears attenuated. This framework has implications for the duration of fast radio bursts and whether they can escape their surroundings.

% \begin{description}
% Secondary publications and information retrieval purposes.
% You may use the \texttt{description} environment to structure your abstract;
% use the optional argument of the \verb+\item+ command to give the category of each item. 
% \end{description}
\end{abstract}

\maketitle

%\tableofcontents

% \section{\label{sec:level1}Introduction}

\textit{Introduction}---Magnetized electron-positron pairs fill the magnetospheres of neutron stars and black holes~\citep{Goldreich&Julian_69, Blandford&Znajek_77} which emit coherent radio signals with the highest brightness temperatures ($\sim10^{33}-10^{41}$\,K) known: giant pulses~\cite{Cognard+96} and fast radio bursts~\citep[FRBs;][]{Lorimer+07, Petroff+19, Cordes&Chatterjee_19, Petroff+22}. Near their sources, such waves are `strong': i.e., the Lorentz-invariant strength parameter $a_0\equiv eE_0/\me c\,\omega_0$ greatly exceeds unity, so electrons oscillate relativistically in the wave field~\cite{Gunn&Ostriker_71, Landau&Lifshitz_75}; here, $E_0$ is the amplitude of the wave electric field, $\omega_0=2\pi c/\lambda_0$ its frequency, and $\me$ and $e$ are the mass and charge of the electron. Whether such waves emerge through the pair rich surroundings constrains where and how they are produced \cite{Beloborodov_21b,Qu+22,Beloborodov_24,Lyutikov_24,Sobacchi+24b,Sobacchi_25,Sridhar+26}.

Besides $a_0$, the interaction of electromagnetic (EM) waves with magnetized pair plasmas depends on three frequencies, defined in the frame where the plasma ahead of the pulse is at rest. These are, $\omo$, the plasma frequency $\omega_{\rm p}\equiv(8\pi n_0e^2/\me)^{1/2}$ ($n_0$ is the proper density of each species), and the cyclotron frequency $\omB\equiv eB_0/m_{\rm e}c$ of the upstream field $B_0$. The plasma magnetization is $\sigma\equiv\omB^2/\omega_{\rm p}^2$. The cold X-mode dispersion relation in pair plasmas allows a subluminal fast-magnetosonic (FMS) branch at $\omo<\omB$ and a superluminal, vacuum-like branch at $\omo^2>\omp^2+\omB^2$, separated by a stop band~\cite{Melrose_97}.

The low-frequency band has been well studied: large-amplitude FMS waves can steepen into shocks \cite{Chen+22a,Lyubarsky_03a,Beloborodov_23}, and decay into Alfv\'en waves through resonant three-wave interactions~\cite{Golbraikh&Lyubarsky_23,Solanki+26b}. For $\omo\ll\omB$, the background field suppresses the particle quiver motion, 
and induced scattering is weakened by gyroradius and Debye-screening effects~\cite{Nishiura+26a,Kamijima+26,Nishiura+26b}; these kinetic theories assume non-relativistic particles in the wave field, $a_0\omo/\omB\ll1$.

On the superluminal branch, $\varepsilonB\equiv a_0\omB/\omo$ separates two regimes. For $1<\varepsilonB<a_0^2$ (i.e., $\omB/a_0<\omo<a_0\omB$) the plasma current is a nonlinear function of the wave field~\cite{Sobacchi+24b} and the pulse can be absorbed by synchrotron absorption~\cite{Lyubarsky_18,Sobacchi_25}, equivalent to the stochastic heating of test particles~\cite{Beloborodov_22,Beloborodov_24}; the exact range of frequencies where this happens is debated. For $\varepsilonB<1$, the current is linear and the particles exhibit relativistic oscillation in the wave field when $a_0>1$~\cite{Sobacchi+24b}. In the latter regime, parametric instabilities such as filamentation~\cite{Sobacchi+22, Sobacchi+23, Iwamoto+23} and induced scattering~\cite{Nishiura+25,Sridhar+26,Iwamoto&Ioka_26}, can still limit wave propagation. In unmagnetized pair plasmas, $\varepsilonp\equiv a_0\omp/\omo$ controls the nonlinearity for all $a_0$: when $\varepsilonp<1$ induced Compton scattering limits the unattenuated pulse to $\simeq\varepsilonp^{-2/3}$ wavelengths; when $\varepsilonp>1$, the head of the pulse launches a shock and gets dissipated~\cite{Sridhar+26}. \textit{What dictates the nature of strong, superluminal EM wave propagation in magnetized plasmas when $\varepsilonB<1$?} This regime applies to FRBs propagating in magnetar winds beyond $\sim10^{12}$\,cm, where $\omo>\omB$ in the magnetar wind-frame and $\varepsilonB<1$~\cite{Sobacchi_25}. Magnetized induced-scattering theories have thus far assumed $\omo<\omB$, weak waves~\cite{Nishiura+25,Nishiura+26a,Nishiura+26b}, or omit plasma response~\cite{Beloborodov_22,Qu&Kumar_26}.

In this \textit{Letter}, we combine analytical theory and kinetic simulations to study monochromatic X-mode pulses propagating along $\hat x$, with electric field oscillating along $\hat y$, across a uniform magnetic field along $\hat z$. This is akin to the geometry of a pulse propagating radially across the toroidal field of a neutron-star wind~\cite{Cerutti&Beloborodov_17}. We derive the induced Compton scattering rate for $a_0>1$; the propagation length is found to depend on $\varepsilonp$, and the ratio of a magnetic restoring frequency $\OmB$ to the scattering rate $\eta$. Kinetic plasma simulations spanning $\sigma=10^{-2}$--$10^{4}$ are performed to compare with our analytical scalings.

\textit{Analytical theory}---We consider a monochromatic, linearly polarized X-mode EM pulse traveling along $+\hat{x}$ in pair plasma. Its oscillating electric field is along $\hat{y}$, its magnetic field is along $\hat{z}$, and so is the uniform background magnetic field, $B_0\hat{z}$. In the upstream plasma rest frame, its frequency and wavenumber are $\omega_0$ and $k_0$. We adopt a cold two-fluid model for the plasma in which the electrons and positrons have the same number densities, the same velocities along $\hat{x}$, and move in the $xy$ plane with opposite velocities along $\hat{y}$. We assume that all physical quantities depend only on $x$ and $t$; taking $c=1$, the governing equations reduce to \cite{Gruzinov_19, Sobacchi+24b}
\begin{align}
\label{eq:main1}
& \frac{\partial}{\partial t}\left(\gamma n\right) + \frac{\partial}{\partial x}\left(nu_x\right) = 0, \\
\label{eq:main2}
& \frac{\partial u_x}{\partial t} + \frac{\partial \gamma}{\partial x} = u_y\frac{n}{n_0} \omB, \\
\label{eq:main3}
& \frac{\partial^2 u_y}{\partial t^2} - \frac{\partial^2 u_y}{\partial x^2} +\omp^2\frac{n}{n_0}u_y = -\omB \left[\frac{\partial}{\partial t}\left( \frac{n u_x}{n_0}\right) + \frac{\partial}{\partial x}\left(\frac{\gamma n}{n_0}\right)\right] \;,
\end{align}
where $n$ is the proper density, $u_x$ and $u_y$ are the $x$ and $y$ components of the particles' four-velocity with a Lorentz factor $\gamma=(1+u_x^2+u_y^2)^{1/2}$.

Because we consider superluminal waves (i.e., $\omo/k_0>1$), we can boost along $\hat{x}$ with a velocity $k_0/\omo$. In this `wave frame', the EM wave vector (and the spatial dependence in the equations) vanishes and the wave oscillates with frequency $\omega=\sqrt{\omega_0^2-k_0^2}$~\cite{Clemmow_74}; however, the spatial dependence of any perturbations remains.  Retaining variations on a timescale $\sim1/\omega$ in the wave frame and linearizing Eqs.~\eqref{eq:main1}-\ref{eq:main3} yields the X-mode dispersion relation \cite{Melrose_97, Sobacchi+24b},
\be
\omega^2 = {\cal A}\omp^2; \ \ \ \ \ {\cal A} \equiv \frac{\omega^2}{\omp^2} = \frac{1}{1-\omB^2/\omo^2}.
\ee
${\cal A}$ ranges from 1 to $1+\sigma$ for the superluminal branch. The unperturbed solution of Eqs.~\eqref{eq:main1}-\eqref{eq:main3} in the wave frame is independent of $x$, and can be written as, $\bar{\gamma} \bar{n}=\gamma_0n_0$, $\bar{u}_x = -u_0 + \chi(t)$, and $\bar{u}_y=\frac{\ahat}{2}(e^{-i\omega t}+e^{i\omega t})$. Here, $u_0=k_0/\omega$, $\gamma_0=\sqrt{1+u_0^2}=\omo/\omega$, $\ahat = {\cal A}a_0$, and $\bar{\gamma}=\sqrt{1 + \bar{u}_x^2 + \bar{u}_y^2}$. Given this, Eq.~\eqref{eq:main2} becomes,
\be
\frac{\partial\chi}{\partial t} = \omB\left(\frac{\gamma_0}{\bar{\gamma}}\right)\bar{u}_y,
\ee
where, 
\be
\label{eq:gamma0gamma}
\dfrac{\gamma_0}{\bar{\gamma}} \approx 1 + \frac{u_0}{\gamma_0}\frac{\chi}{\gamma_0} + \frac{1}{2}\left(\frac{3u_0^2}{\gamma_0^2} - 1 \right)\frac{\chi^2}{\gamma_0^2} - \frac{\bar{u}_y^2}{2\gamma_0^2} + \cdots
\ee
This yields the leading order solution of $\chi(t) = \frac{\omB}{\omega}\ahat\sin{\omega t}$.
Eq.~\eqref{eq:main3} becomes,
\be
\frac{\partial^2 \bar{u}_y}{\partial t^2}+\omp^2\frac{\gamma_0}{\bar{\gamma}}\bar{u}_y = -\omB\gamma_0\frac{\partial}{\partial t}\left( \frac{-u_0+\chi}{\bar{\gamma}}\right) \;.
\ee

We introduce the following small perturbations to the solutions to study the stability of the unperturbed solution: $\gamma n=\gamma_0n_0+\delta\rho$, $u_x = \bar{u}_x + \delta u_x$, and $u_y = \bar{u}_y + \delta u_y$. Inserting this in Eq.~\eqref{eq:main1}-\ref{eq:main3} and upon linearizing, we get,
\begin{align}
\label{eq:pert1}
& \frac{\partial}{\partial t} \left(\frac{\delta\rho}{\gamma_0n_0}\right) + \frac{\partial}{\partial x}\left[\frac{1}{\gamma_0}\delta\left(\frac{nu_x}{n_0}\right)\right] = 0,\\
\label{eq:pert2}
& \frac{\partial \delta u_x}{\partial t} + \frac{\partial \delta\gamma}{\partial x} = \omB\delta\left(\frac{nu_y}{n_0}\right),\\
\label{eq:pert3}
& \frac{\partial^2\delta u_y}{\partial t^2} - \frac{\partial^2\delta u_y}{\partial x^2} + \omp^2 \delta\left(\frac{nu_y}{n_0}\right)  = -\omB\left[ \frac{\partial}{\partial t}\delta\left(\frac{nu_x}{n_0}\right) + \frac{\partial}{\partial x}\frac{\delta\rho}{n_0} \right], \;
\end{align}
where,
\begin{align}
\label{eq:pert1_approx}
& \delta\left(\frac{n u_x}{n_0}\right) = \gamma_0\left[ \frac{\bar{u}_x}{\bar{\gamma}}\frac{\delta \rho}{\gamma_0n_0} + \frac{(1+\bar{u}_y^2)\delta u_x - \bar{u}_x\bar{u}_y\delta u_y}{\bar{\gamma}^3} \right]\nonumber\\
& \approx \gamma_0\left[-v_x\frac{\delta \rho}{\gamma_0 n_0} + \frac{2(1-v_x)}{\gamma_0}\delta u_x + \frac{v_x\bar{u}_y}{\gamma_0^2}\delta u_y\right],
\end{align}
\begin{align}
\label{eq:pert2_approx}
& \delta\left(\frac{nu_y}{n_0}\right) = \gamma_0\left[ \frac{\bar{u}_y}{\bar{\gamma}}\frac{\delta \rho}{\gamma_0n_0} + \frac{(1 + \bar{u}_x^2)\delta u_y - \bar{u}_x\bar{u}_y\delta u_x}{\bar{\gamma}^3}\right]\ \nonumber\\
& \approx \frac{\gamma_0^3}{\bar{\gamma}^3}\left(1-\frac{2u_0\chi}{\gamma_0^2} + \frac{\chi^2}{\gamma_0^2}\right)\delta u_y + \bar{u}_y\frac{\gamma_0}{\bar{\gamma}}\left(\frac{\delta \rho}{\gamma_0 n_0} + \frac{v_x}{\bar{\gamma}}\delta u_x\right),\\
\label{eq:pert3_approx}
&\delta \gamma = \frac{\bar{u}_x \delta u_x + \bar{u}_y\delta u_y}{\bar{\gamma}} \approx -v_x\delta u_x + \frac{\bar{u}_y}{\gamma_0}\delta u_y.
\end{align}

In order to determine $v_x$, let us first expand $\bar{u}_x/\bar{\gamma}=$
\be
-\left[1+\frac{1+\bar{u}_y^2}{(u_0-\chi)^2}\right]^{-1/2} \approx -1+\frac{1+\bar{u}_y^2}{2(u_0-\chi)^2} + {\cal O}\left(\frac{(1+\bar{u}_y^2)^2}{(u_0-\chi)^4}\right)
\ee
and expand
\be
\label{eq:u0-chi}
\frac{1}{(u_0-\chi)^2} \approx \frac{1}{\gamma_0^2}\left[1+2\frac{\chi}{\gamma_0}+3\frac{\chi^2}{\gamma_0^2}+\cdots\right].
\ee
Then, we can take $\langle\bar{u}_y^2\rangle\approx\ahat^2/2$, $\langle\chi\rangle=0$, $\langle\chi^2\rangle/\gamma_0^2=({\cal A}\varepsilonB)^2/2$, $\langle \chi (1+\bar{u}_y^2)\rangle=0$, $\langle(1+\bar{u}_y^2)\chi^2\rangle=(\ahat\omB/\omega)^2\left(\frac{1}{2}+\frac{\ahat^2}{8}\right)$ and $\gamma_0\simeq u_0$, to obtain the following phase-averaged quantity,
\be
\label{eq:baruxgamma}
 \left\langle\frac{\bar{u}_x}{\bar{\gamma}}\right\rangle \approx -v_x + {\cal O}\left[\frac{1+\ahat^2/2}{\gamma_0^2}({\cal A}\varepsilonB)^2 + \left(\frac{1+\ahat^2/2}{\gamma_0^2}\right)^2\right],
\ee
from which we see that $v_x=\left(1 - \dfrac{1 + \ahat^2/2}{2\gamma_0^2}\right)$; the correction terms in the following quantity,
\be
\left\langle\frac{\gamma_0^3}{\bar{\gamma}^3}\left(1-\frac{2u_0\chi}{\gamma_0^2} + \frac{\chi^2}{\gamma_0^2}\right)\right\rangle = 1 + \frac{3u_0^2-\gamma_0^2}{2\gamma_0^4}\langle\chi^2\rangle - \frac{3\langle\bar{u}_y^2\rangle}{2\gamma_0^2} + \cdots
\ee
are of the order ${\cal O}\left[({\cal A}\varepsilonB)^2 + \ahat^2/\gamma_0^2\right]\ll1$, and can be ignored. From Eqs.~\eqref{eq:u0-chi}-\ref{eq:baruxgamma}, we see that the fast-timescale-averaged longitudinal velocity changes ($\chi$-dependent terms) are suppressed in a relativistic mean motion in the regime of $({\cal A}\varepsilonB)^2\ll 1$, thus allowing us to neglect the longitudinal correction term (${\cal O}(\chi^2/\gamma_0^2)$) while retaining the transverse correction term (${\cal O}(\ahat^2/\gamma_0^2)$). Also, since $\bar{u}_y/\bar{\gamma} = \frac{\bar{u}_y}{\gamma_0}(1+u_0\chi/\gamma_0^2)+\cdots$ (where $\chi \bar{u}_y=(\ahat^2\omB/2\omega)\sin{2\omega t}$) does not alter the leading coefficient at frequency $\omega$, we consider $\bar{u}_y/\bar{\gamma}\simeq\bar{u}_y/\gamma_0$. Taking into account the approximations mentioned above and substituting Eqs.~\eqref{eq:pert1_approx}-\ref{eq:pert3_approx} into Eqs.~\eqref{eq:pert1}-\ref{eq:pert3}, we get the following continuity equation:
\be
\label{eq:continuity}
(\partial_t - v_x\partial_x)\left(\frac{\delta \rho}{\gamma_0n_0}\right) + \frac{2(1-v_x)}{\gamma_0}\partial_x\delta u_x + \frac{v_x\bar{u}_y}{\gamma_0^2}\partial_x\delta u_y=0,
\ee
longitudinal momentum equation:
\be
\label{eq:long_mom}
(\partial_t - v_x\partial_x)\delta u_x + \frac{\bar{u}_y}{\gamma_0}\partial_x\delta u_y =
\omB\left[\delta u_y + \bar{u}_y\left(\frac{\delta \rho}{\gamma_0n_0} + \frac{v_x}{\gamma_0}\delta u_x\right)\right],
\ee
and the transverse wave equation:
\begin{align}
\label{eq:trans_mom}
    & (\partial_t^2-\partial_x^2)\delta u_y + \omp^2\left[\delta u_y + \bar{u}_y\left(\frac{\delta \rho}{\gamma_0n_0} + \frac{v_x}{\gamma_0}\delta u_x\right)\right] = \nonumber \\
    & -\omB\gamma_0\bigg[-v_x\partial_t\frac{\delta \rho}{\gamma_0n_0} + \frac{2(1-v_x)}{\gamma_0}\partial_t\delta u_x \nonumber\\
    &+\frac{v_x}{\gamma_0^2}\left(\dot{\bar{u}}_y\delta u_y + \bar{u}_y\partial_t\delta u_y\right) + \partial_x\left(\frac{\delta \rho}{\gamma_0n_0}\right)\bigg].
\end{align}

The periodic coefficients of Eqs.~\eqref{eq:continuity}-\eqref{eq:trans_mom} couple to perturbations having different frequencies. We look for solutions of the form, 
\begin{align}
\delta\rho/\gamma_0 n_0 &= e^{i(kx - \Omega t)}\left[Q_0 + Q_+ e^{-i\omega t} + Q_- e^{i\omega t} + \cdots\right] \nonumber \\
\delta u_x &= e^{i(kx - \Omega t)}\left[U_0 + U_+ e^{-i\omega t} + U_- e^{i\omega t} + \cdots\right] \nonumber \\
\delta u_y &= e^{i(kx - \Omega t)}\left[A_0 + A_+ e^{-i\omega t} + A_- e^{i\omega t} + \cdots\right].
\end{align} 
Substituting these expressions into Eqs.~\eqref{eq:continuity}-\eqref{eq:trans_mom} and using Eqs.~\eqref{eq:pert1_approx}-\eqref{eq:pert3_approx}, we can isolate the coefficients corresponding to $e^{-i\Omega t}$ to obtain the following equations:
\begin{align}
D_{\rm v} Q_0 & = \frac{2(1-v_x)}{\gamma_0} kU_0 + \frac{   \ahat v_x}{2\gamma_0^2} k \left(A_+ + A_-\right) \label{eq:four1}, \\
D_{\rm v} U_0 & = \frac{\ahat}{2\gamma_0} k \left(A_+ + A_-\right)  + i\omB \left(A_0 + \frac{\ahat}{2}\Sigma\right). \label{eq:four2}
\end{align}
Comparing the above two equations with Eqs.~\eqref{eq:pert1}-\ref{eq:pert2}, we get $\delta(nu_x/n_0)|_\Omega=\gamma_0\Omega Q_0/k$ and $\delta(nu_y/n_0)|_\Omega = A_0 + \ahat\Sigma/2$, where $\Sigma\equiv Q_+ + Q_- + v_x(U_+ + U_-)/\gamma_0$. Substituting these in Eq.~\eqref{eq:pert3} yields,
\be \label{eq:four3}
(k^2 - \Omega^2 + \omp^2)A_0 = -i\omB\gamma_0\frac{k^2-\Omega^2}{k}Q_0 - \frac{\omp^2\ahat}{2}\Sigma.
\ee
In a similar way, we can isolate the coefficients corresponding to $e^{-i(\Omega\pm\omega)t}$ to obtain the following continuity and longitudinal momentum equations:
\begin{align}
(D_{\rm v}\pm\omega)Q_\pm & = \frac{2(1-v_x)k}{\gamma_0}U_\pm + \frac{v_x\ahat k}{2\gamma_0^2}A_0,
\label{eq:four4}\\
(D_{\rm v}\pm\omega)U_\pm & = \frac{\ahat k}{2\gamma_0}A_0 + i\omB\left[A_\pm + \frac{\ahat}{2}\left(Q_0 + \frac{v_xU_0}{\gamma_0}\right)\right],
\label{eq:four5}
\end{align}
where $D_{\rm v} \equiv \Omega + v_x k$.
Comparing the above two equations with Eqs.~\eqref{eq:pert1}-\ref{eq:pert2}, we get $\delta(nu_x/n_0)|_{(\Omega\pm\omega)}=\gamma_0(\Omega\pm\omega) Q_\pm/k$ and $\delta(nu_y/n_0)|_{(\Omega\pm\omega)} = A_\pm + \ahat(Q_0+v_xU_0/\gamma_0)/2$. Substituting these in Eq.~\eqref{eq:pert3} yields,
\be \label{eq:DAplusminus}
D_\pm A_\pm & = - \dfrac{i\omB\gamma_0}{k}(D_\pm + \omp^2)Q_\pm + \dfrac{\omp^2\ahat}{2} \left(Q_0 + \dfrac{v_xU_0}{\gamma_0} \right),
\ee
where
\be
D_\pm & \equiv \left(\Omega\pm \omega \right)^2 -k^2-\omp^2.
\ee
Eqs.~\eqref{eq:four1}-\ref{eq:DAplusminus} constitute a system of nine equations and nine amplitude variables; requiring the corresponding coefficient matrix to have a zero determinant determines the dispersion relation. Using the above equations, we can solve for the coefficients, 
\begin{align}
    A_0 & = -i\omB\gamma_0\frac{Q_0}{k}f_0 - \frac{\ahat}{2}(1-f_0)\Sigma; \label{eq:A0} \\ 
    A_\pm & = \frac{\omp^2\ahat}{2D_\pm} \left(Q_0+\frac{v_xU_0}{\gamma_0}\right) - \frac{i\omB\gamma_0}{k}\frac{(D_\pm+\omp^2)}{D_\pm}Q_\pm; \\
    A_+ + A_- & = \frac{\omp^2\ahat}{2}\left(Q_0 + \frac{v_xU_0}{\gamma_0}\right)\left(\frac{1}{D_+} + \frac{1}{D_-}\right) \nonumber\\
    & - \frac{i\omB\gamma_0}{k}\left[\frac{D_++\omp^2}{D_+}Q_+ + \frac{D_-+\omp^2}{D_-}Q_-\right].
\end{align}
We can eliminate $A_0$ by inserting Eq.~\eqref{eq:A0} into Eq.~\eqref{eq:four2}; the resultant equation can then be substituted into a form of Eq.~\eqref{eq:four1} multiplied by $D_{\rm v}$. This yields,
\begin{align}
(D_{\rm v}^2 - f_0\OmB^2)Q_0 & = \frac{\ahat k}{2\gamma_0^2}\left[2(1-v_x)k + v_x D_{\rm v}\right](A_+ + A_-) \nonumber \\ 
& + \frac{ik\omB\ahat f_0}{\gamma_0}(1-v_x)\Sigma.
\end{align}
Likewise, after eliminating the other amplitude variables, we obtain the following dispersion relation,
\begin{equation}
\label{eq:DR}
D_{\rm v}^2 - f_0\OmB^2 \simeq \frac{2\eta^3}{\omega} k\left(\Omega+k\right)\left(\frac{1}{D_+}+\frac{1}{D_-}\right).
\end{equation}
Here, we define the magnetic restoring frequency, $\OmB \equiv \sqrt{2(1-v_x)}\omB = \sqrt{1+\ahat^2/2}(\omB/\gamma_0)$, the scattering rate $\eta\equiv\omega({\cal A\varepsilonp}/2)^{2/3}$, and $f_0\equiv\frac{k^2-\Omega^2}{k^2-\Omega^2+\omp^2} \rightarrow 1$ in the limit $|k^2-\Omega^2|\gg\omp^2$. In the following, we will consider that $|D_{\rm v}|\ll|\Omega + k|$. 

Substituting the resonance conditions $k=k_{\rm m}$ and $\Omega=\Omega_{\rm m}+\Delta\Omega$ into Eq.~\eqref{eq:DR} yields the maximum growth rate of the instability.  Here, $k_{\rm m}$ and $\Delta\Omega$ are determined by the conditions (i) $D_{\rm v}=0$ and $D_\pm=0$ when $\OmB\ll\eta$ (strong coupling), which gives $\Omega_{\rm m}=-v_xk_{\rm m}$, and (ii) $D_{\rm v}^2=\OmB^2$ and $D_\pm=0$ when $\OmB\gg\eta$ (weak coupling), which gives $\Omega_m=-v_xk_m-\OmB$. For the $D_{\rm v}\pm\OmB=0$ branches, this yields, 
\be \label{eq:eta3}
\Delta \Omega^2(\Delta\Omega\mp2\OmB) = \eta^3. 
\ee
But, one can see by perturbing $D_{\rm v}=-\OmB + \Delta\Omega$ that only the $D_{\rm v}=-\OmB$ solution to $\Delta\Omega$ yields an imaginary value for all $\OmB$, and has the largest growth rate. Note that the aforementioned conditions $D_+=0$ and $D_-=0$ do not hold simultaneously. The most unstable wave number is,
\begin{align} \label{eq:Kstar}
k_{\rm m} =
    \begin{cases}
          - \dfrac{2\gamma_0^2\omega}{1+\ahat^2/2} & (\OmB\ll\eta)\\
          - \dfrac{2\gamma_0^2\omega}{1+\ahat^2/2}\left(1-\dfrac{\OmB}{\omega}\right) & (\OmB\gg\eta) \\
    \end{cases}
    ,
\end{align}
and the growth rate of the instability, ${\rm Im}(\Delta\Omega)$, is,
\begin{align} \label{eq:Ostar}
\Gamma_{\rm gr} =
    \begin{cases}
          \dfrac{\sqrt{3}}{2}\eta & (\OmB\ll\eta)\\
          \dfrac{\eta^{3/2}}{\sqrt{2\OmB}} & (\OmB\gg\eta).\\
    \end{cases}
\end{align}
Eqs.~\eqref{eq:Kstar}-\eqref{eq:Ostar} are valid for any $a_0$ provided $\ahat/\gamma_0={\cal A}^{3/2}\varepsilonp\ll 1$, $v_x\approx1$, ${\cal A}\varepsilonB\ll 1$, and $\OmB\ll\omega$.

From the head of the EM pulse, the distance where induced scattering significantly affects the pulse profile (in the wave frame) is approximately $1/\Gamma_{\rm gr}$. In the reference frame where the plasma ahead of the pulse is at rest, the corresponding distance is,
\begin{align} \label{eq:mainresult}
\ell_{\rm e}/\lambda_0 \simeq \frac{\omega_0/\gamma_0}{2\pi\Gamma_{\rm gr}}\simeq
    \begin{cases}
        \dfrac{2^{2/3}}{\pi\sqrt{3}}({\cal A}\varepsilonp)^{-2/3}  & (\OmB\ll\eta)\\
        \dfrac{\sqrt{2\OmB/\omega}}{\pi{\cal A}\varepsilonp} \xrightarrow[]{\ahat\gg1} \dfrac{2^{1/4}}{\pi\sqrt{{\cal A}}}\dfrac{\sqrt{\varepsilonB}}{\varepsilonp} & (\OmB\gg\eta).\\
    \end{cases}
\end{align}
Eq.~\eqref{eq:mainresult} represents the e-fold growth length for the strong and weak coupling regimes. The strong and weak coupling regimes meet when $\OmB=\eta$ at a critical frequency of $(\omo/\omB)_\star=\sqrt{1+\ahat^2/2}({\cal A}\varepsilonp/2)^{-2/3}$. Beyond the linear propagation length from its head, the EM pulse is attenuated because the energy of the forward-propagating pump wave is transferred into compressing the background magnetic fields, the scattered waves, and plasma density/velocity fluctuations. Ref.~\cite{Lyutikov_26} derives exact nonlinear dispersion relations for strong circularly polarized waves propagating along a background magnetic field, but not induced-scattering rates, so a direct comparison with our results is not possible. Further, as a consistency check, Eqs.~\eqref{eq:Kstar}-\ref{eq:mainresult} reduce to Eqs.~(14)-(16) of Ref.~\cite{Sridhar+26} in the limit of $\omB\rightarrow0$.

\begin{figure*}
    \centering
    \includegraphics[width=0.33\linewidth]{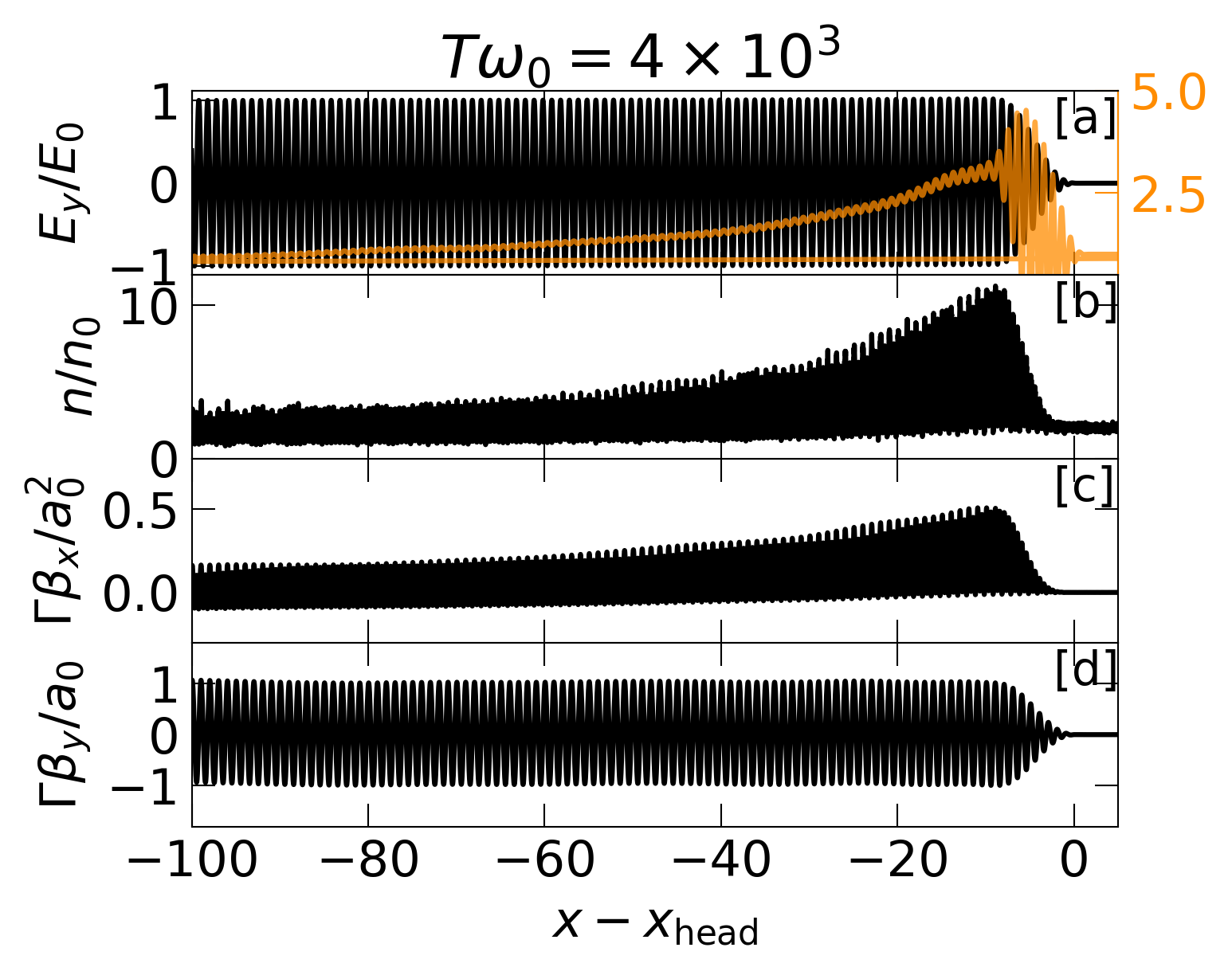}
    \includegraphics[width=0.31\linewidth]{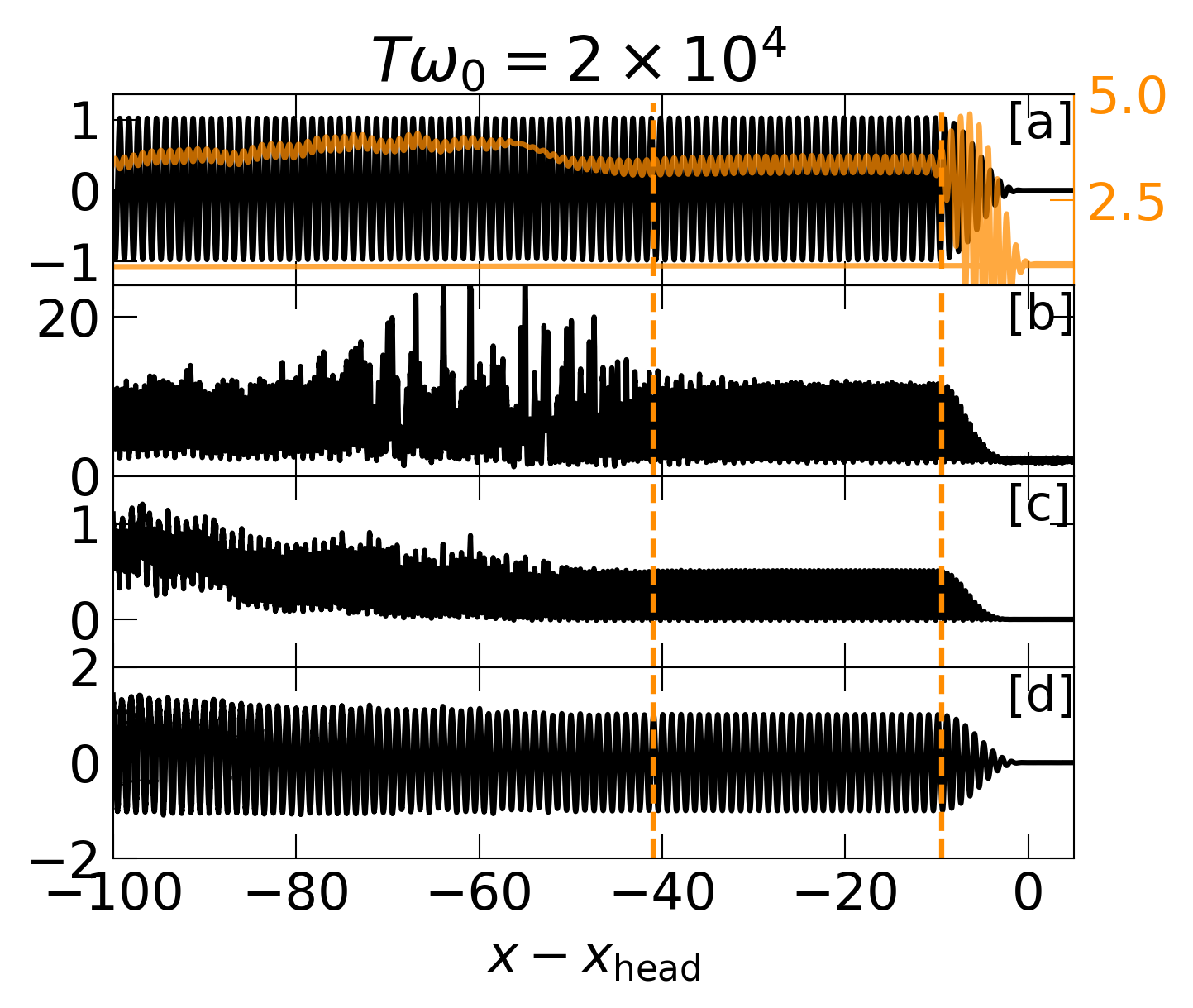}
    \includegraphics[width=0.33\linewidth]{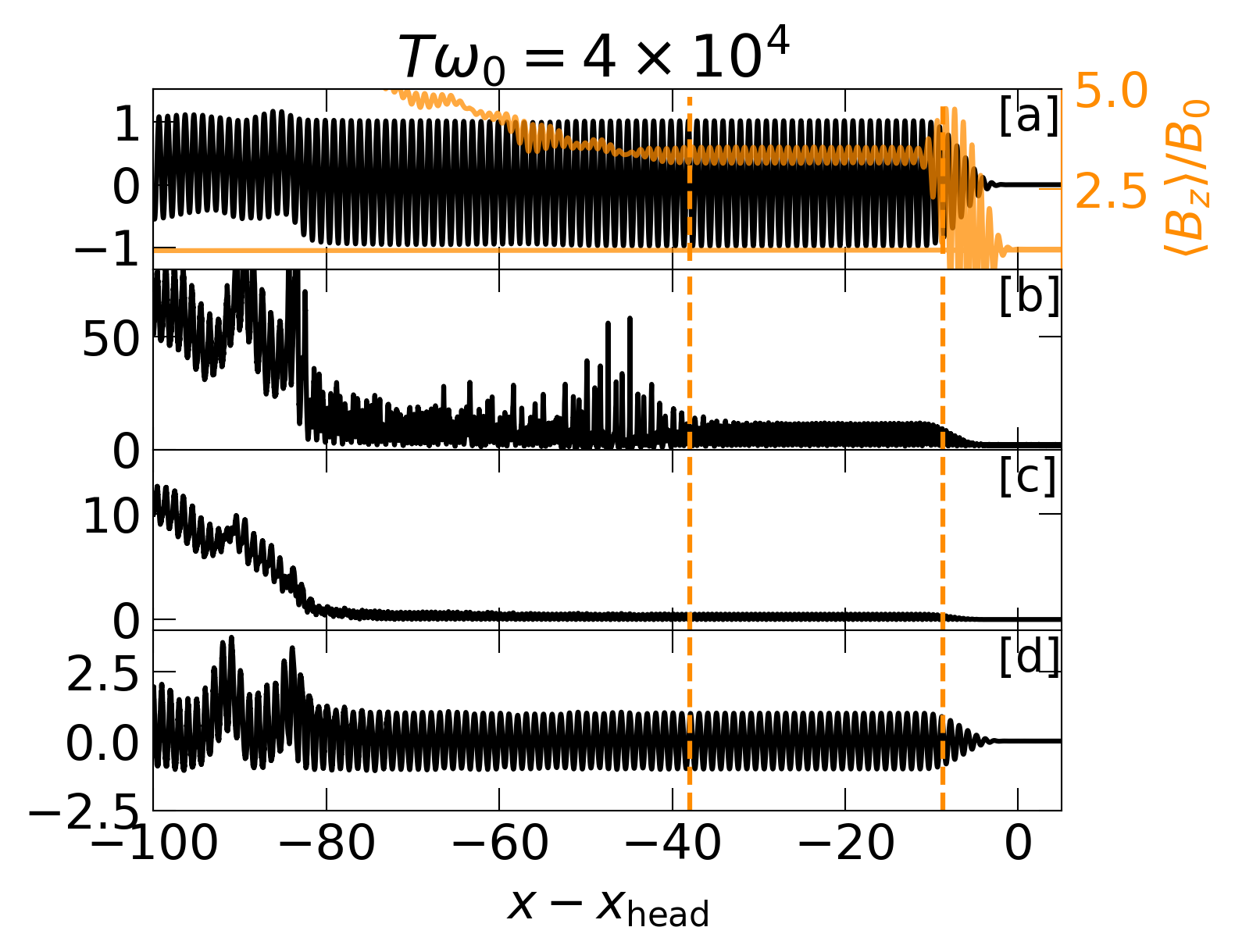}
    \caption{Snapshots of the EM pulse-plasma interaction region for a simulation with $a_0=3$, $\sigma=10$ and $\omo/\omp=100$, i.e., in the relativistic/linear regime where $\omo>a_0\omB$. Different columns represent different stages of evolution (the entire pulse is in vacuum at time $T\omo=0$). [a] The wave's electric field (black), and phase-averaged magnetic field (orange) normalized by their respective incident-wave amplitudes, [b] simulation-frame particle number density in units of the upstream density, and cell-averaged four-velocity of electrons [c] along the $\hat{x}$ direction (normalized by $a_0^2$) and [d] along the $\hat{y}$ direction (normalized by $a_0$). The orange dashed vertical lines in the middle and right columns enclose the linear propagation length, $\llinear$. The distance from the head of the EM pulse, $x-x_{\rm head}$, is expressed in units of the vacuum wavelength.}
    \label{fig:relativistic_linear}
\end{figure*}

\begin{figure*}
    \centering
    \includegraphics[width=1\linewidth]{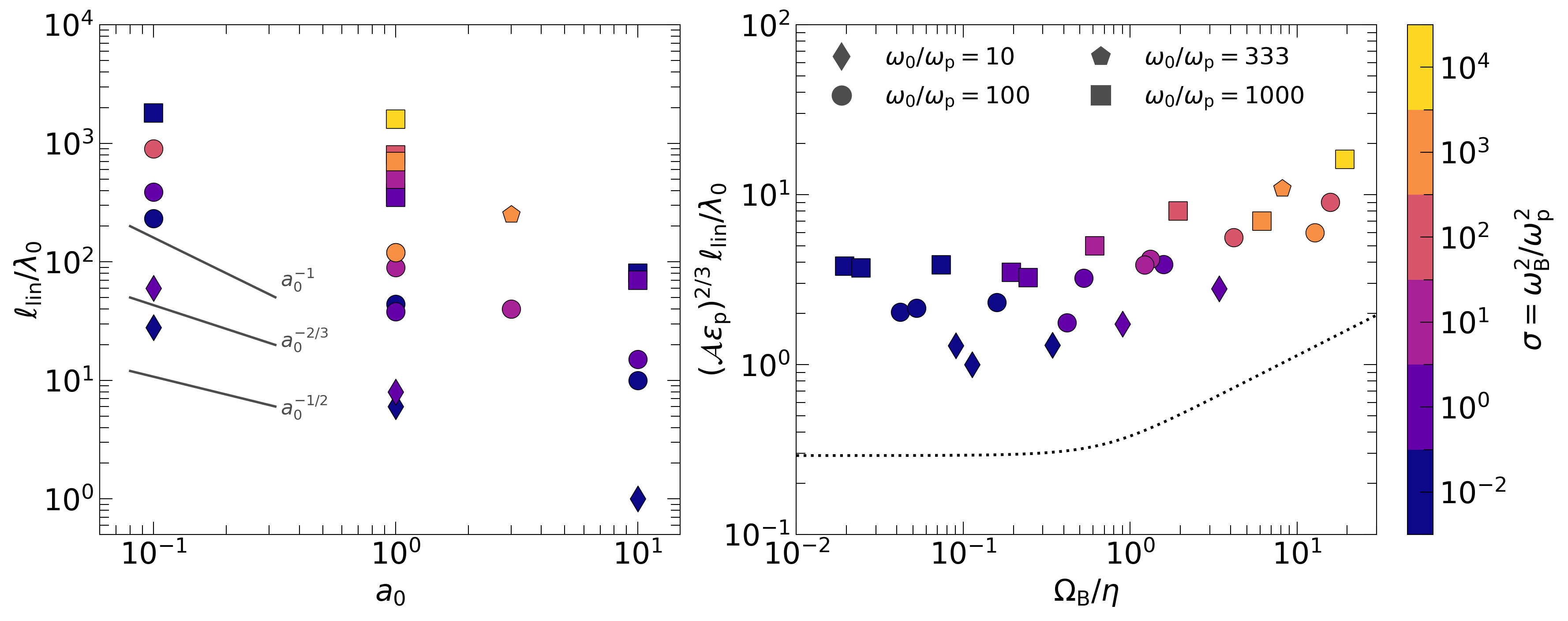}
    \caption{Left: Linear propagation length in units of the vacuum wavelength as a function of $a_0$; grey line-segments denote reference slopes. Right: Linear propagation length in units of the vacuum wavelength, multiplied by $({\cal A}\varepsilonp)^{2/3}$, as a function of $\OmB/\eta$ (Eq.~\eqref{eq:mainresult}). The markers denote numerical simulation results (belonging to the linear current regime with $\omo>a_0\omB$ for $a_0>1$); their shape and color represent $\omo/\omp$ and $\sigma$, respectively. The black dotted curve denotes the one e-fold analytical scaling across the strong ($\OmB<\eta$) and weak coupling ($\OmB>\eta$) regimes (see Eq.~\eqref{eq:eta3}).}
    \label{fig:lin_prop_length}
    \vspace{-0.4cm}
\end{figure*}

\textit{Numerical simulations}---We perform fully kinetic particle-in-cell (PIC) simulations with {\tt OSIRIS} \cite{Fonseca+02}. Our effectively one-dimensional setup follows that of Ref.~\cite{Sridhar+26}, but with an added uniform field $B_0\hat z$ throughout the computational domain. We launch an EM pulse from vacuum along $+\hat{x}$ with fields $E_y\hat{y}$ and $B_z\hat{z}$, hence an X-mode. 
Its amplitude gradually rises over 10-100 wavelengths to its maximum value (corresponding to $a_0$), and stays flat for long enough to capture nonlinear effects. The cold plasma is initially at rest in the simulation frame. It enters the computational domain through a density ramp at the right boundary, and leaves from the left edge, while the window follows the EM pulse along $\hat{x}$ at $c$ (more in the End Matter). We vary $a_0=[0.1,1,3,10]$, $\omop=[10,100,333,1000]$, and $\sigma=[10^{-2},1,10,10^2,10^3,10^4]$, retaining only combinations above the superluminal X-mode cutoff $\omo/\omp>\sqrt{1+\sigma}$.

\textit{Flux compression}---We find that the background magnetic flux is frozen-in with the plasma, which is compressed by the EM pulse in both the linear and the nonlinear regimes. For a plasma moving at $\sim c$, the phase-averaged quantities yield $\langle B_z\rangle-\langle E_y\rangle=B_0$. The plasma drifting at $\beta_{\rm D}$ introduces a motional electric field $\langle E_y\rangle\simeq\beta_{\rm D}\langle B_z\rangle$. In the steady state, the background magnetic field in the pulse is compressed by a factor $\langle B_z\rangle/B_0\simeq1+\ahat^2/4$. We show an example of this with the orange curve in the top row of Fig.~\ref{fig:relativistic_linear}: within the linear head of the pulse, $\langle B_z\rangle/B_0$ saturates over time to a value of $\sim3.3$, close to $1+\ahat^2/4=3.25$ for the simulated $a_0=3$. At a given distance from the head of the pulse, the saturation is reached after a time $\propto\sigma$, because the frozen-in field adds inertia to the plasma~\cite{Beloborodov_26}. The steady compression is $\mathcal O(\ahat^2)$ in the upstream frame but only $\mathcal O(\ahat^2/\gamma_0^2)$ in the wave frame (Eqs.~36-44 of Ref.~\cite{Sobacchi+24b}). Therefore, induced scattering growth rate and $\ell_{\rm e}$ only weakly depend on whether the $\sigma$-dependent relaxation has finished. %$t_\star\omega_0\sim2\pi(1+\sigma)(1+a_0^2/4)\,\xi/\lambda_0$

\textit{The linear regime}---When $\varepsilonB<1$, we verify that the particles' velocity is relativistic and the plasma current is linear in the wave field \cite{Sobacchi_25}. In Fig.~\ref{fig:relativistic_linear}, we present an example of this case with $a_0=3$, $\sigma=10$ and $\omo/\omp=100$ at different stages of interaction. At later stages of the simulation (middle and right columns), the vertical orange dashed lines enclose and highlight the linear portion of the interaction, which evolves only modestly with time once identified. In all our simulation runs with $\varepsilonB<1$, the initialized wave propagates through the plasma, albeit for a finite length; the activation of induced Compton scattering introduces large amplitude fluctuations in density and momenta a distance behind the pulse head, and inhibits propagation of the trailing part of the pulse.

\textit{Linear propagation length}---We follow the same procedure as Ref.~\cite{Sridhar+26} to measure the linear propagation length, $\llinear={\cal N}\ell_{\rm e}$, which is a multiple of the instability e-fold length, based on the departure of $\Gamma\beta_{\rm y}$ from $eE_y/\me c \omo$. 
The left panel of Fig.~\ref{fig:lin_prop_length} shows the measured $\llinear/\lambda_0$ against $a_0$: at fixed $(\sigma,\omo/\omp)$, Eq.~\eqref{eq:mainresult} predicts that $\llinear/\lambda_0\propto a_0^{-1}$ (for $\ahat\ll1$) and $\propto a_0^{-1/2}$ (for $\ahat\gg1$) in the weak coupling regime ($\OmB\gg\eta$), and $\llinear\propto a_0^{-2/3}$ in the strong coupling regime ($\OmB\ll\eta$). These slopes are similar, making it challenging to identify this transition from the measurements (left panel). On the other hand, Eq.~\eqref{eq:mainresult} also implies that $({\cal A}\varepsilonp)^{2/3}\llinear/\lambda_0$ depends on $\OmB/\eta$ alone. We test this in the right panel of Fig.~\ref{fig:lin_prop_length}: the colored markers denote simulation measurements and the black dotted curve is a solution to Eq.~\eqref{eq:eta3} for its complex root with Im($\Delta\Omega)>0$ as a function of $\OmB/\eta$. In our PIC simulations, we measure the distance when the instability becomes visible above intrinsic particle noise, $\ell_{\rm lin}$. From Fig.~\ref{fig:lin_prop_length}, we measure $\ell_{\rm lin}/\ell_{\rm e}$ to be ${\cal N}\sim4$ e-folds. We note that the numerical measurements of $({\cal A}\varepsilonp)^{2/3}\llinear/\lambda_0$ are consistent with our analytical scaling across six decades in $\sigma$, following a flat slope for $\OmB\ll\eta$ as in the unmagnetized case \cite{Sridhar+26} and rising as $(\OmB/\eta)^{1/2}$ for $\OmB\gg\eta$.

\begin{figure*}
    \centering
    \includegraphics[width=0.33\linewidth]{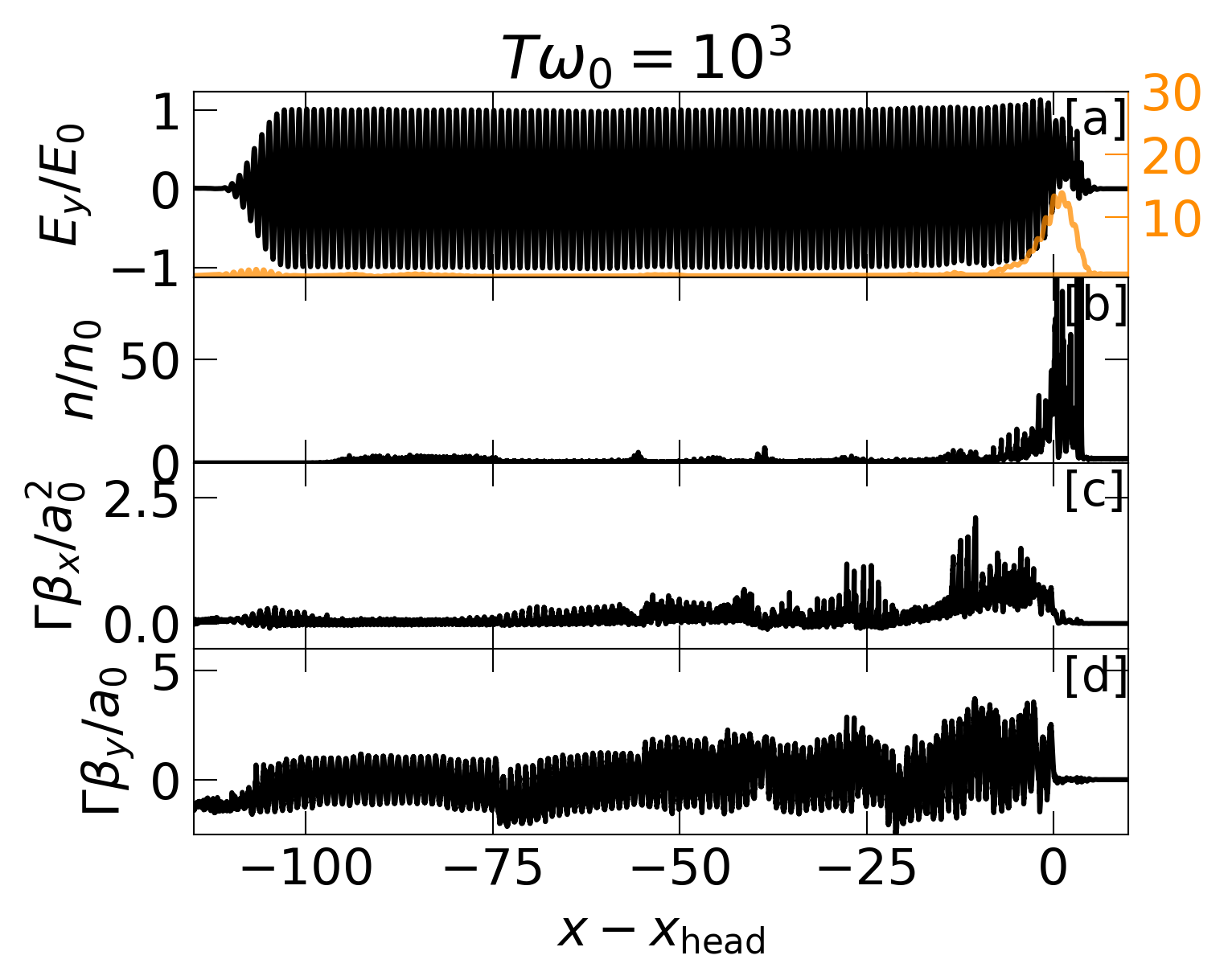}
    \includegraphics[width=0.31\linewidth]{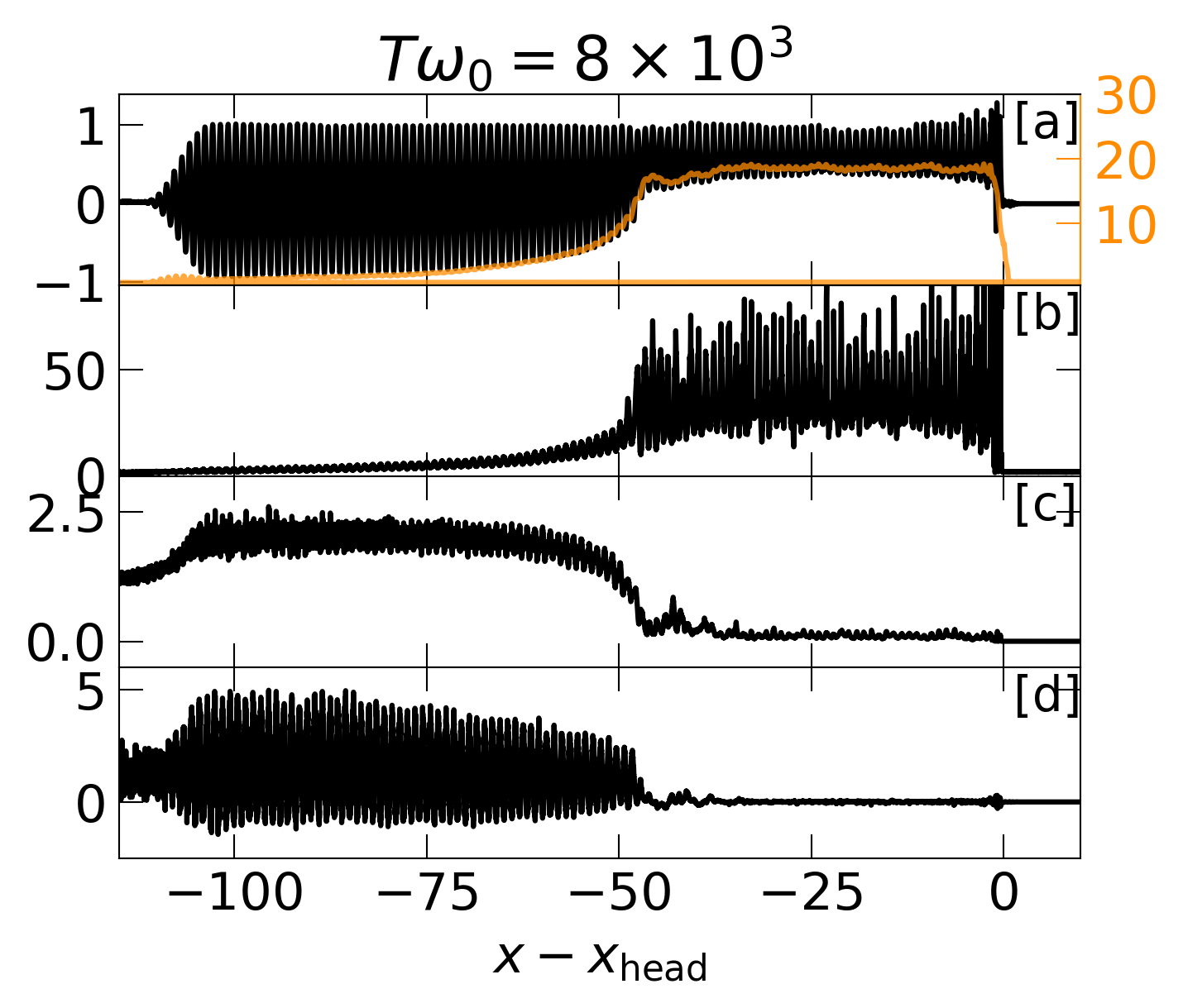}
    \includegraphics[width=0.33\linewidth]{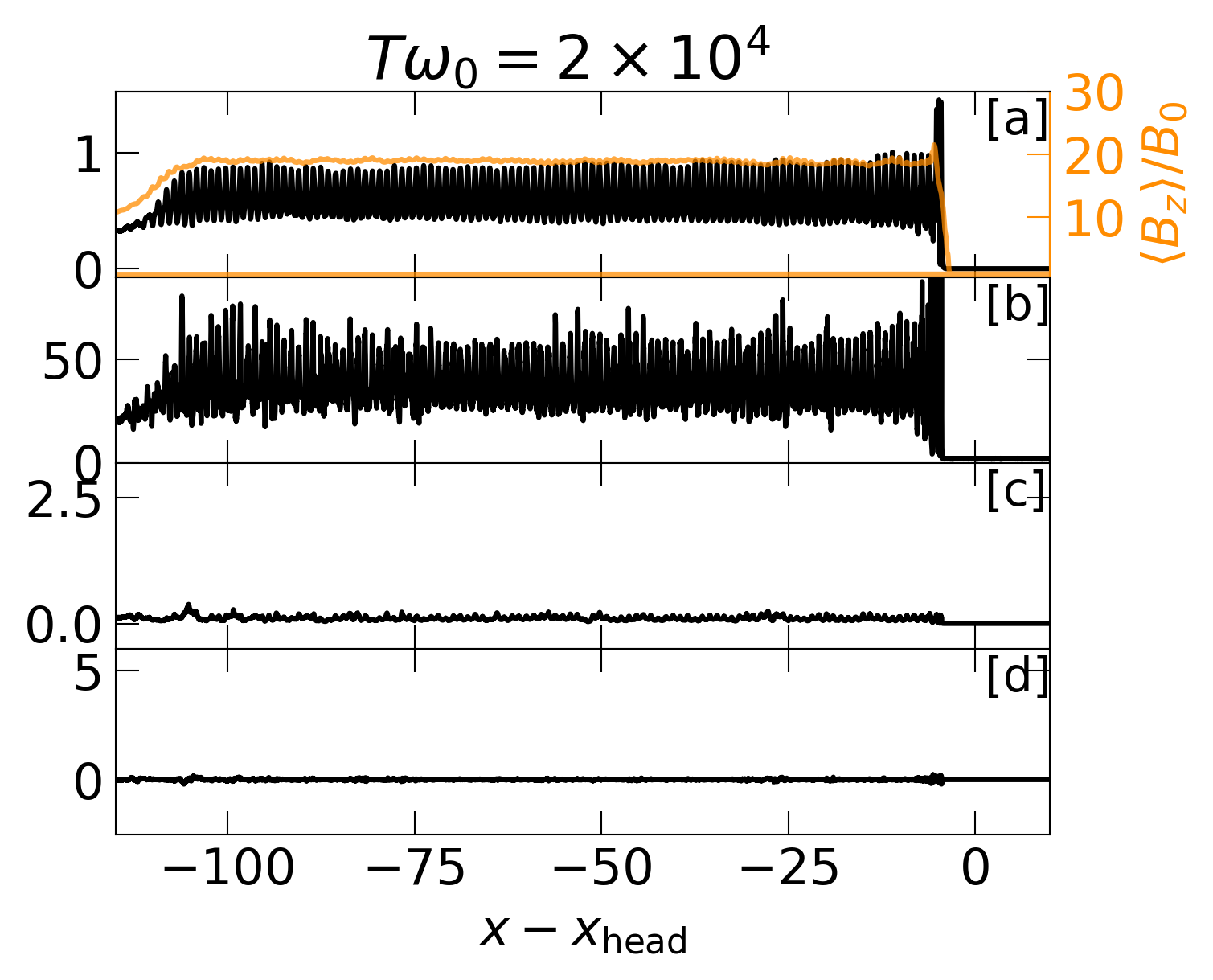}
    \caption{Snapshots of the EM pulse-plasma interaction region for a simulation with $a_0=10$, $\sigma=10$ and $\omo/\omp=10$, i.e., in the relativistic/non-linear regime where $a_0^2>\varepsilonB>1$. Different rows [a-d] denote the same quantities as in Fig.~\ref{fig:relativistic_linear}.}
    \label{fig:relativistic_nonlinear}
    \vspace{-0.4cm}
\end{figure*}

\textit{The nonlinear regime}---We identify in our PIC simulations, the particles' velocity to be relativistic and their response to the wave field to be nonlinear when $a_0^2>\varepsilonB>1$~\cite{Sobacchi+24b}.  In Fig.~\ref{fig:relativistic_nonlinear}, we show the spatial structure of the EM pulse-plasma interaction region (for an example case with $a_0=10$, $\sigma=10$ and $\omo/\omB\approx3.2$) at different stages of interaction (columns). 
Behind the head of the pulse, the plasma settles into a compression front~\cite{Beloborodov_26}. The particles exhibit $\mathbf E\times\mathbf B$ drift along $\hat x$ in this compressed field rather than a coherent quiver motion (mean transverse momentum $\Gamma\beta_y\simeq0$, row [d]). As discussed previously, $E_y$ acquires a motional component due to the drifting plasma on which the oscillation rides. Empirically, we find that the amplitude of the oscillations is damped to a value $\simeq E_0/\varepsilonB$, where the local resonance condition of $a\simeq\omo/\omB$ switches on~\cite{Sobacchi_25}. We interpret the apparent attenuation of the half-cycle with $E_y<0$ as a damped oscillation riding on a positive motional field, and not as an attenuation of one phase. Unlike previous works~\cite{Chen+22a}, this wave does not steepen into strong shocks in each wavelength. For a vacuum-like wave with electric (and magnetic) field amplitudes $E_{\rm w}$ propagating on these mean fields, $B^2-E^2=(\langle B_z\rangle+\langle E_y\rangle+2E_{\rm w})(\langle B_z\rangle-\langle E_y\rangle)$, where $\langle B_z \rangle - \langle E_y\rangle=B_0$ is the upstream value. Without field compression, phases with $E_{\rm w}<-(\langle B_z \rangle + \langle E_y\rangle)/2=-B_0/2$ violate the MHD condition $B^2>E^2$ and dissipate at shocks in each wavelength~\cite{Chen+22a, Beloborodov_23}. 
Once the compressed flux attains $\langle B_z \rangle/B_0\simeq1+\ahat^2/4$, $\langle B_z\rangle+\langle E_y\rangle$ increases to $(1+\ahat^2/2)B_0$. Therefore, the condition for not producing electric-dominated shocks becomes $2E_0/\varepsilonB\lesssim(1+\ahat^2/2)B_0$, so the wave in Fig.~\ref{fig:relativistic_nonlinear} is absorbed resonantly rather than through strong shocks at each wave cycle.

\textit{Conclusions}---We have studied, analytically and with fully kinetic simulations, the interaction of strong ($a_0>1$) linearly polarized superluminal X-mode EM waves with magnetized pair plasmas. The interaction is nonlinear when $a_0\gtrsim1$ in electron--proton plasmas~\cite{Kruer_19} and when $\varepsilonp\gtrsim1$ in unmagnetized pair plasmas~\cite{Sobacchi+24a,Sridhar+26}. In magnetized pair plasmas ($\sigma>1$), this regime separation is set by the parameter $\varepsilonB\equiv a_0\omB/\omo$.

(1)~For $\varepsilonB<1$ the EM pulse propagates until induced Compton scattering disrupts it. The background field imparts an effective restoring frequency to the instability at $\OmB$, which competes with the wave scattering rate $\eta$. For $\OmB\ll\eta$, the instability proceeds as in unmagnetized plasmas, and the linear propagation distance in units of the pulse wavelength is $\llinear/\lambda_0\propto({\cal A}\varepsilonp)^{-2/3}$. For $\OmB\gg\eta$ the instability growth slows and $\llinear/\lambda_0$ increases to $\propto(\OmB/\omega)^{1/2}/({\cal A}\varepsilonp)$, which tends to $\sqrt{\varepsilonB}/\varepsilonp$ for $\ahat\gg1$. Kinetic plasma simulations spanning $\sigma=10^{-2}$--$10^4$ and $a_0=0.1$--10 confirm the predicted analytical scalings.

(2)~For $a_0^2>\varepsilonB>1$, the plasma current is nonlinear~\cite{Sobacchi+24b}. Behind the head of the pulse, the plasma relaxes into a compression front~\cite{Beloborodov_26} where the particles exhibit a longitudinal $\mathbf E\times\mathbf B$ drift with subdued transverse momenta. In the presented case, flux compression ensures strong shocks are not formed at each wave. The EM pulse rides on top of a motional electric field and is gradually damped, consistent with synchrotron absorption~\cite{Lyubarsky_18, Beloborodov_22, Sobacchi_25} (so one half-cycle appears attenuated).

(3)~In both the regimes, the EM pulse compresses the background field to $\langle B_z\rangle/B_0\simeq1+\ahat^2/4$ over a time $\propto\sigma$~\cite{Beloborodov_26}. The compression is only ${\cal O}(\ahat^2/\gamma_0^2)$ in the wave frame \cite{Sobacchi+24b}, so $\llinear$ is weakly sensitive to it.

These results could be applicable to large amplitude, linearly polarized radio pulses such as FRBs and giant radio pulses, which eventually cross the predominantly toroidal field of a neutron star's wind. There, the wind-frame cyclotron frequency falls below GHz frequencies outside $\sim10^{11}$\,cm, and $\varepsilonB<1$ beyond $\sim10^{12}$\,cm~\cite{Sobacchi_25}. Since $\sigma\sim10^2$--$10^3$ in such winds~\cite{Beloborodov_26}, $\varepsilonp=\varepsilonB/\sigma^{1/2}\ll1$. For $\ahat\gg1$, $(\OmB/\eta)^3\simeq\sqrt2\,{\cal A}\sigma\varepsilonB$, so induced scattering there is in the weak-coupling regime wherever $\sigma\varepsilonB\gtrsim1$. There, magnetization lengthens $\llinear$ relative to the unmagnetized estimate by a factor of $\approx1.2(\OmB/\eta)^{1/2}$, i.e., 2--$4\times$ for $\sigma\varepsilonB\sim10$–$10^3$.
The predictions of this theory could be tested with next-generation multi-petawatt laser and accelerator experiments~\cite{Chen&Fiuza_23, Arrowsmith+24} capable of producing dense pair plasmas. Broadband pulses~\cite{Lyubarsky_08,Nishiura+26a}, multidimensional, and radiative effects remain to be studied.

\textit{Acknowledgments---}This work benefits from discussions with Roger D. Blandford, Lorenzo Sironi, and Emanuele Sobacchi. N.S. acknowledges the support of the Simons Foundation (grant MP-SCMPS-00001470) and the National Aeronautics and Space Administration (grant 80NSSC26K0946). The computing for this project was performed on Stanford University's \textit{Sherlock} cluster. Anthropic's Claude was used to assist with plotting in Matplotlib \cite{Hunter_07} and checking algebra, under author's supervision. N.S. thanks Stanford University and their Research Computing Center for providing computational resources and support that contributed to this research.

\textit{Data availability---}The data supporting the findings of this article are not publicly available upon publication because it is not technically feasible and/or the cost of preparing and hosting the data would be prohibitive within the terms of this research project. The data are available from the author upon reasonable request.

% \nocite{*}

\bibliography{aps}% Produces the bibliography via BibTeX.

\cleardoublepage
% \appendix

\section{End Matter}
% \subsection{Numerical simulation details} \label{sec:numerical_details}

\textit{Numerical simulation details---}Our numerical simulations use the Higuera \& Cary particle pusher \citep{Higuera&Cary_17}, and the shape of the macroparticles is represented with quadratic splines. 
The EM field solver employs a second-order finite-difference method using an extended 4-coefficient stencil along the wave-propagation direction ($\hat{x}$), which improves dispersion errors. This choice requires the computational domain to be two-dimensional, and the quadratic particle splines require at least 5 cells along the $\hat{y}$ direction.
The simulation setup overall follows Ref.~\cite{Sridhar+26}; we repeat it here for completeness. At $t=0$ the EM pulse is initialized in a vacuum region, located to the left of the space filled with plasma. The entire domain is initialized with a uniform magnetic field along the $\hat{z}$ direction. The amplitude of the EM pulse is gradually increased and decreased over a length of $\ell_{\rm rise} = \ell_{\rm fall} = 10-100$ wavelengths using a smooth 5th-order polynomial profile function of the form $f(x) = 10l(x)^3 - 15l(x)^4 + 6l(x)^5$, where $l(x) = (x - x_{\rm head})/\ell_{\rm rise}$ for the rising phase of the pulse (right) and $l(x) =(x_{\rm head} + \ell_{\rm flat} - x)/\ell_{\rm fall}$ for the falling phase (left). Here, $x_{\rm head}$ is the position of the right-edge of the pulse, and $\ell_{\rm flat}=10^2-10^4$ wavelengths is length of the `flat' part of the pulse (with strength parameter $a_0$), which is chosen to be sufficiently long to capture nonlinear propagation effects. The computational domain along $\hat{x}$ is chosen to encompass the entire pulse.  We resolve the wavelength of the EM pulse with $\ge 100$ cells, the electron skin depth $c/\omp$ with $160-16000$ cells, and both the cold Larmor radius and the Debye length $\lambda_{\rm D}=\sqrt{\theta_0}(c/\omp)$ with at least one cell for our cold plasma with a normalized temperature of $\theta_0=10^{-4}$. Our simulations employ 50 particles per cell (equally divided between electrons and positrons).
In the transition zone between the vacuum and the plasma, the density profile smoothly increases from 0 to 1 using a hyperbolic tangent function of the form $g(x) = 0.5\left[1 + \tanh{(x-1.1 x_{\rm head})}\right]$. The plasma entering the domain through the right boundary exit through an absorbing boundary modeled as a perfectly matched layer \cite{Vay_02} on the left.
The choice of a smooth profile in the amplitude of the EM pulse and in the plasma density results in a gradual increase of the current and minimizes potential numerical artifacts during the early stages of interaction. The plasma is initially at rest, and the simulation window follows the pulse as it propagates through the plasma.

%\textit{Analytical details---}

\end{document}